\documentclass{emulateapj}

\usepackage{graphicx,amsmath,booktabs,threeparttable,url}
\usepackage{hyperref}

\usepackage{threeparttable}

\usepackage[usenames,dvips]{color}

\usepackage{natbib}
\begin{document}

\title{Impact of Type Ia supernova ejecta on a helium-star binary companion}

\author{Kuo-Chuan Pan$^1$, Paul M. Ricker$^1$ and Ronald E. Taam$^{2,3}$}
\affil{$^1$Department of Astronomy, University of Illinois at Urbana-Champaign, 1002 W. Green Street, Urbana, IL 61801; kpan2@illinois.edu, pmricker@illinois.edu}
\affil{$^2$Department of Physics and Astronomy, Northwestern University, 2131 Tech Drive, Evanston, IL 60208; taam@tonic.astro.northwestern.edu}
\affil{$^3$Academia Sinica Institute of Astronomy and Astrophysics, P.O. Box 23-141, Taipei 10617, Taiwan}


\begin{abstract}
The impact of Type Ia supernova ejecta on a helium-star companion is investigated via high-resolution,
two-dimensional hydrodynamic simulations. For a range of helium-star models and initial binary
separations it is found that the mass unbound in the interaction, $\delta M_{\rm ub}$, is related
to the initial binary separation, $a$, by a power law of the form $\delta M_{\rm ub} \propto
a^{m}$. This power-law index is found to vary from $-3.1$ to $-4.0$, depending on the 
mass of the helium star. The small range of this index brackets values found previously for 
hydrogen-rich companions, suggesting that the dependence of the unbound mass on orbital separation 
is not strongly sensitive to the nature of the binary companion. 
The kick velocity is also related to the initial binary separation by a power law with an index 
in a range from $-2.7$ to $-3.3$, but the power-law index differs from those found in previous studies 
for hydrogen-rich companions. 
The space motion of the companion after the supernova is dominated by its 
orbital velocity in the pre-supernova binary system. The level of Ni/Fe contamination of the
companion resulting from the passage of the supernova ejecta is difficult to estimate, but an 
upper limit on the mass of bound nickel is found to be $\sim 5\times 10^{-4}\ M_\odot$.
\end{abstract}

\keywords{binaries: close --- supernovae: general --- methods: numerical}

\section{INTRODUCTION}

Type Ia supernovae (SNe~Ia) are among the most catastrophic and energetic events in the Universe.
Because they are so luminous, and because their
light-curve shapes and absolute magnitudes are correlated, SNe~Ia can be used as
``standardizable candles''
in measuring the distances to remote galaxies, allowing constraints to be placed on
key cosmological parameters \citep{Branch:1992p1603, Riess:1996p1623}.
SNe~Ia also play a major role in galactic chemical evolution via their energy and metal input
to the interstellar medium.
Thus, the nature of their progenitor systems and the physical origin of variations
in their properties are fundamental problems of great interest \citep{Dominguez:2001p2574,
Howell:2007p2557}.

Most work on progenitor models for SNe~Ia has focused on two general classes of systems.
The single-degenerate scenario involves a CO white dwarf (WD) accreting matter from a non-compact 
stellar binary companion, eventually becoming unstable to explosive nuclear burning 
\citep{Whelan:1973p78,Nomoto:1982p1467}. On the other hand, the double-degenerate scenario involves 
the merger of two CO WDs whose orbital decay results from the loss of angular momentum due to 
gravitational wave emission \citep{Iben:1984p43, Webbink:1984p104}.

Single-degenerate models require that the rate of mass accretion be such that the WD avoid a nova
explosion.  For a high accumulation efficiency, stable burning on the WD is required, but this burning
phase occurs in a fairly narrow range above $10^{-7}\ M_\odot\ {\rm yr}^{-1}$.  For rates greater 
than this range, a wind from the WD develops that limits the accretion efficiency 
\citep{Nomoto:1982p1467,Hachisu:1996p3228,Ivanova:2004p123}, making it difficult to explain the 
observed low-redshift SN~Ia rate of $3 \times 10^{-5}{\rm \ Mpc}^{-3}{\rm\ yr}^{-1}$ 
\citep{Mannucci:2005p3244} using single-degenerate progenitors alone. Although the 
predicted and observed numbers of double-degenerate systems are sufficient to explain the observed 
rate \citep{Nelemans:2001p3347,Napiwotzki:2001p3317, Napiwotzki:2002p3341}, it is more likely that 
WD-WD mergers lead to the production of an ONeMg WD followed by accretion-induced collapse to a neutron 
star \citep{Nomoto:1985p3367, Ivanova:2004p123,Dessart:2006p3379, Wickramasinghe:2009p3391}.  Moreover, 
the wide range of total WD masses suggests that double-degenerate models should exhibit much more 
heterogeneity than is observed in SN~Ia light curves \citep{Goldhaber:2001p3408,Knop:2003p3425}.

\cite{Maoz:2008p3181} analyzed the observed SN~Ia rate for assumed initial mass functions (IMF)
and concluded that almost all intermediate-mass close binary systems in the range $3 -8\ M_{\sun}$
should evolve to the SN~Ia stage. This result allows for a wide range of possibilities for SN~Ia progenitor 
systems. Furthermore, observations of the SN~Ia rate as a function of redshift suggest the need for a
two-component model for the delay time distribution (DTD). (The delay time is the interval between a
star's arrival on the zero-age main sequence and its destruction in an SN~Ia.)
\cite{Scannapieco:2005p2469} 
and \cite{Mannucci:2006p2431} found that the observations can be fit with a short-delay-time population having
delays of $\sim 10^8$~yr and a long-delay-time population having delays of $3 - 4$~Gyr.

To provide an explanation for the two populations suggested by the observed
SN~Ia rate, several pre-supernova progenitor models have been investigated.  The long-delay-time
population can be understood in terms of progenitor systems characterized by a main-sequence-like 
companion in the MS-WD channel \citep{Hachisu:2008p144} and/or by a red giant in the RG-WD channel
\citep{Hachisu:1996p3228, Hachisu:1999p3434, Hachisu:1999p3439,Hachisu:2008p3845}. In contrast, the 
short-delay-time population may consist of systems characterized by a massive MS star in the MS-WD 
channel or by a He star in the He-WD channel \citep{Waldman:2008p628, Wang:2009p3601, Wang:2009p3585,
Wang:2009p3598,Meng:2009p3843}.

Numerically, \cite{Marietta:2000p112} explored the influence of the supernova explosion on the 
companion star in the single-degenerate channel with hydrogen-rich stars consisting of red giants, 
subgiants, and main-sequence stars using two-dimensional Eulerian hydrodynamics simulations. They 
found that significant quantities of hydrogen would be unbound from the companion star in each case 
(15\% of the envelope for main-sequence and subgiant cases, and 98\% of the red giant envelope), in 
conflict with observational upper limits on the amount of hydrogen inferred from SN~Ia spectra 
\citep{Mattila:2005p1442,Leonard:2007p102}.

A more recent hydrodynamical study by \cite{Pakmor:2008p139} reexamined the main-sequence 
simulation of  \cite{Marietta:2000p112} using a three-dimensional smoothed particle hydrodynamics (SPH) simulation.  
In contrast to Marietta et al., Pakmor et al.\ adopted the structure for the companion star based on the binary 
evolutionary models of \cite{Ivanova:2004p123}, which yielded more compact main-sequence-like 
companions.  As a consequence, Pakmor et al.\ found a tenfold reduction in the amount of unbound mass 
compared with Marietta et al., bringing the prediction of the amount of unbound hydrogen-rich material
into agreement with the observational upper limits.

In a complementary analytical study, \cite{Kasen:2009p2965} investigated the radiation emitted by the
collision of SN~Ia ejecta with a red giant, finding that the light curve should depend on the viewing angle.
This result was attributed to the fact that the gas is more transparent in the region 
shadowed by the companion star. This suggests the possibility that the secondary star may be detectable in
future observational studies.

These previous simulations of the effect of a supernova impact on a
companion star have been carried out only for models applicable to the long-delay-time population of SN~Ia. 
In contrast, \cite{Wang:2009p3601, Wang:2009p3585} suggest a progenitor binary model based upon a 
helium-star channel for the short-delay-time population.  Using a binary evolution model, for this channel
they find a SN~Ia birthrate $\sim 3 \times 10^{-4}$~yr$^{-1}$ and a corresponding delay time of
$\sim 4.5 \times 10^7$~yr to $\sim 1.4 \times 10^8$~yr
\citep{Wang:2009p3598}. The latter delay time is consistent with that estimated from observations
for the short-delay-time population by \cite{Scannapieco:2005p2469}, \cite{Mannucci:2006p2431}, and 
\cite{Aubourg:2008p2503}. 

In this paper we report the results of Eulerian hydrodynamics simulations of the impact of SN~Ia ejecta on
companion stars for the single-degenerate helium-star channel.
In the next  section, the assumptions underlying our study, the construction of the initial model, and the numerical 
method are described.  Our numerical results for a range of helium-star models
and orbital separations are reported in \S~3 and their implications are discussed in \S~4. 
In the final section, we summarize our results and make some concluding remarks.  

\section{NUMERICAL METHODS AND MODELS}

\subsection{Numerical codes}
For our hydrodynamical simulations we used FLASH version 3 \citep{Fryxell:2000p109,dubey_introduction_2008}. FLASH is a parallel, 
multi-dimensional hydrodynamics code based on block-structured adaptive mesh refinement (AMR).  To solve the
Euler equations on the AMR grid, we used the piecewise parabolic method 
\citep{Colella:1984p25} with modifications to handle nonideal equations of state \citep{Colella:1985p66}.
The equation of state used
is interpolated from a precomputed table of the Helmholtz free energy. It includes 
contributions from radiation, completely ionized nuclei, and degenerate electrons and positrons 
\citep{Timmes:2000p1} for an optically thick mixture of gas and radiation in local 
thermodynamic equilibrium.   

\begin{figure}
\plotone{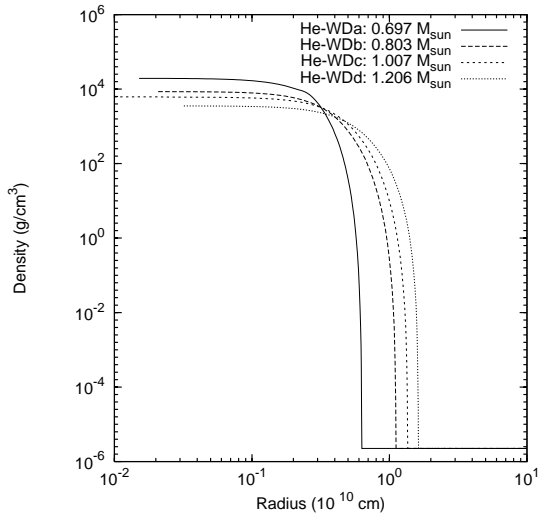}
\caption{\label{dens} Density profiles of the helium-star models at the onset of the supernova explosion,
as obtained using the one-dimensional stellar evolution code. }
\end{figure}

The helium-star models used in our simulations were generated using a one-dimensional stellar evolution code 
\citep{Eggleton:1971p89,Eggleton:1972p114,Eggleton:1973p3453}. 
We simulated four helium-star models 
with initial masses equal to 1.25, 1.35, 1.4, and $1.8\ M_\sun$. To evolve the helium-star models 
to the onset of the supernova explosion, an artificial constant mass loss rate was adopted such that the 
evolution time and final helium star masses were consistent with the detailed binary evolutionary 
models of \cite{Wang:2009p3585}. The resulting models are summarized in Table~\ref{tab1}. 
The density profiles of the helium-star models at the onset 
of the supernova explosion are illustrated in Figure~\ref{dens}.  It can be seen that the more massive 
models are characterized by larger radii and less compact cores.

\begin{table*}
\begin{center}

\caption{ Helium-star models. \label{tab1}}
\begin{tabular}{cccccccc} 
\\
\hline
\hline
            & Mass & Radius  & Evolution time  & m$_{\rm ub}$\tablenotemark{a} & C$_{\rm ub}$\tablenotemark{a} & m$_{\rm kick}$\tablenotemark{b} &  C$_{\rm kick}$\tablenotemark{b} \\   
Model  & ($M_\sun$) & ($10^{10}$ cm) & (yr)  & & & & \\
\tableline
He-WDa & 0.697 & 0.63 & $9.2 \times 10^{6}$ & -4.01 & 1.42 & -3.28  & 4150\\
He-WDb & 0.803 & 1.10 & $5.2 \times 10^{5}$ & -3.13 & 0.70 & -2.90  & 2413\\
He-WDc & 1.007 & 1.35 & $2.2 \times 10^{5}$ & -3.48 & 1.17 & -3.18 &  2703\\
He-WDd & 1.206 & 1.63 & $6.1 \times 10^{4}$ & -3.51 & 1.35 & -2.71 &  1729\\
\tableline
\end{tabular} \\
$^a$The entries for the power-law index, m$_{\rm ub}$, and power-law constant, C$_{\rm ub}$, refer to\\
the power-law fit to unbound mass versus orbital separation described in eq.~\ref{eq_stripped}. \\
$^b$The entries for the power-law index, m$_{\rm kick}$, and power-law constant, C$_{\rm kick}$, refer\\ to 
the power-law fit to kick velocity versus orbital separation described in eq.~\ref{eq_kick}.
\end{center}
\end{table*}

\subsection{Initial setup}
Since the speed of the ejecta in a SN~Ia ($\lesssim 10^4$~km~s$^{-1}$) is much higher than the orbital speed 
of the helium star ($\lesssim 10^3$~km~s$^{-1}$) in a binary system, we ignore the orbital motion in 
the first approximation and consider a 2D axisymmetric geometry.  The simulation domain is described 
using cylindrical coordinates ($r, z$), with the $z$-axis defined as the direction along the line 
connecting the centers of the white dwarf and the helium star. We consider a simulation domain with a 
size equal to fifteen times the radius of the helium star ($R_{\rm He}$) in the radial direction and 
$30\ R_{\rm He}$ in the axial direction. For convenience, the helium star is located at the origin of the 
coordinate grid. 

To simplify the problem, the composition of the one-dimensional helium-star model was taken to be
a uniform distribution of $98\%$ helium and $2\%$ carbon by mass when used in FLASH.  This 
simplification leads to an error of $\lesssim$~2\% in the composition and $\lesssim$~4\% in 
the radius for the lowest-mass helium-star model. To initialize the two-dimensional FLASH simulations, we first interpolated the one-dimensional model onto the FLASH grid using up to twelve levels of refinement based on the magnitudes of the second derivatives of gas density and pressure. With each block containing $8\times 16$ zones, the equivalent 
uniform-grid resolution is thus $16,384 \times 32,768$.  For model He-WDc, the minimum zone spacing at this
level of refinement corresponds to $1.22 \times 10^7$~cm. Within the helium star, a minimum of nine levels
of refinement was used (corresponding to a maximum zone spacing of $9.76 \times 10^7$~cm for model He-WDc).
Because the surface of the helium star is characterized by a very steep density gradient, we established
a sharp cutoff radius to avoid pressure errors at the surface. The cutoff radius was chosen such that each
density drop of one order of magnitude is resolved by at least three zones in the surface region
(we have more than 40 zones per order of magnitude in the core region). Outside the cutoff radius,
the density was set to an ambient value of $2.25 \times 10^{-6}$~g~cm$^{-3}$, and the pressure was set to
the value of the helium-star model pressure profile at the cutoff radius. The simulations, therefore,
employed the same ambient density but somewhat different ambient pressures. The composition of the ambient
gas in each case was taken to be pure hydrogen.

The helium-star model was relaxed on the Eulerian grid by artificially damping the momentum for a period 
of time greater than about thirty times the average dynamical time scale. During this time the damping 
factor was smoothly increased from $0.7$ to $0.99$, ensuring that the Mach number in the helium-star 
interior was always smaller than $0.01$. Once this process was complete, the damping was removed and the 
gas velocity was reset to zero. The supernova explosion was then introduced to the grid.
During the subsequent evolution, we allowed second-derivative refinement up to seven levels everywhere
except in two regions: within the helium star at any time, and in a region surrounding the supernova
explosion for the first 150 seconds, we required nine levels of refinement (equivalent to a
resolution of $2048 \times 4096$ for a uniform grid with a minimum zone spacing of $9.77 \times 10^7$~cm in  
model He-WDc).  The extra refinement for the supernova region reduces the influence of grid artifacts on
the developing explosion. The explosion itself was introduced by creating a spherical region of high-density
and high-temperature gas with radius equal to twenty times the minimum zone spacing at nine levels of
refinement. Each run used this ``7/9'' refinement pattern except for one model that was also run at
6/8, 8/10, 9/11, and 10/12 to study convergence (see \S~\ref{convergence}).

The Type Ia supernova model used is the W7 model described by \cite{Nomoto:1984p82}, which corresponds 
to a carbon deflagration in an accreting CO WD. The initial carbon deflagration is assumed to develop 
in the central regions of the white dwarf and is then followed by a detonation.
This one-dimensional model provides a good fit to observed light curves and can be approximated by a white 
dwarf of mass $M_{\rm wd} = 1.378\ M_\sun$, total explosion energy $E_{\rm sn}=1.233 \times 10^{51}$~erg, and average 
speed $v_{\rm sn}=8.527 \times 10^3$~km~s$^{-1}$.  This mass is uniformly distributed within the spherical
perturbation used to start the calculation. The velocity inside the perturbation is taken to be
radially outward and uniform in magnitude; the internal energy is set using $v_{\rm sn}$ and $E_{\rm sn}$.  
We assume the ejecta to be entirely $^{56}$Ni and use this fluid
component as a tracer for the ejecta.
The mass of the supernova ejecta creates a potential perturbation that changes the equilibrium state of
the helium star. However, in all of our models the radius of the helium star is smaller than the 
Roche lobe radius, and the timescale on which the equilibrium is upset by the supernova shock,
$\sim a/v_{\rm sn}$ ($a$ is the initial binary separation), is shorter than the average
dynamical timescale in the helium star. Thus we expect that the altered pre-supernova potential should
not have a significant effect on the structure of our models. However, because the orbital 
motion is ignored in this study, the gravitational force from the supernova material will attract 
the helium star and cause a small velocity perturbation toward the supernova material. This velocity 
perturbation can be comparable to the kick velocity of the helium star after supernova impact 
but is much smaller than the ejecta speed.

\section{RESULTS}

In this section, we describe the numerical results for the standard case, model He-WDc,
and explore the dependence of the system's evolution on the mass of the helium-star companion
and the initial binary orbital separation $a$. To determine the sensitivity of the results to the numerical 
resolution, we also describe a convergence test.

\begin{figure*}
\plotone{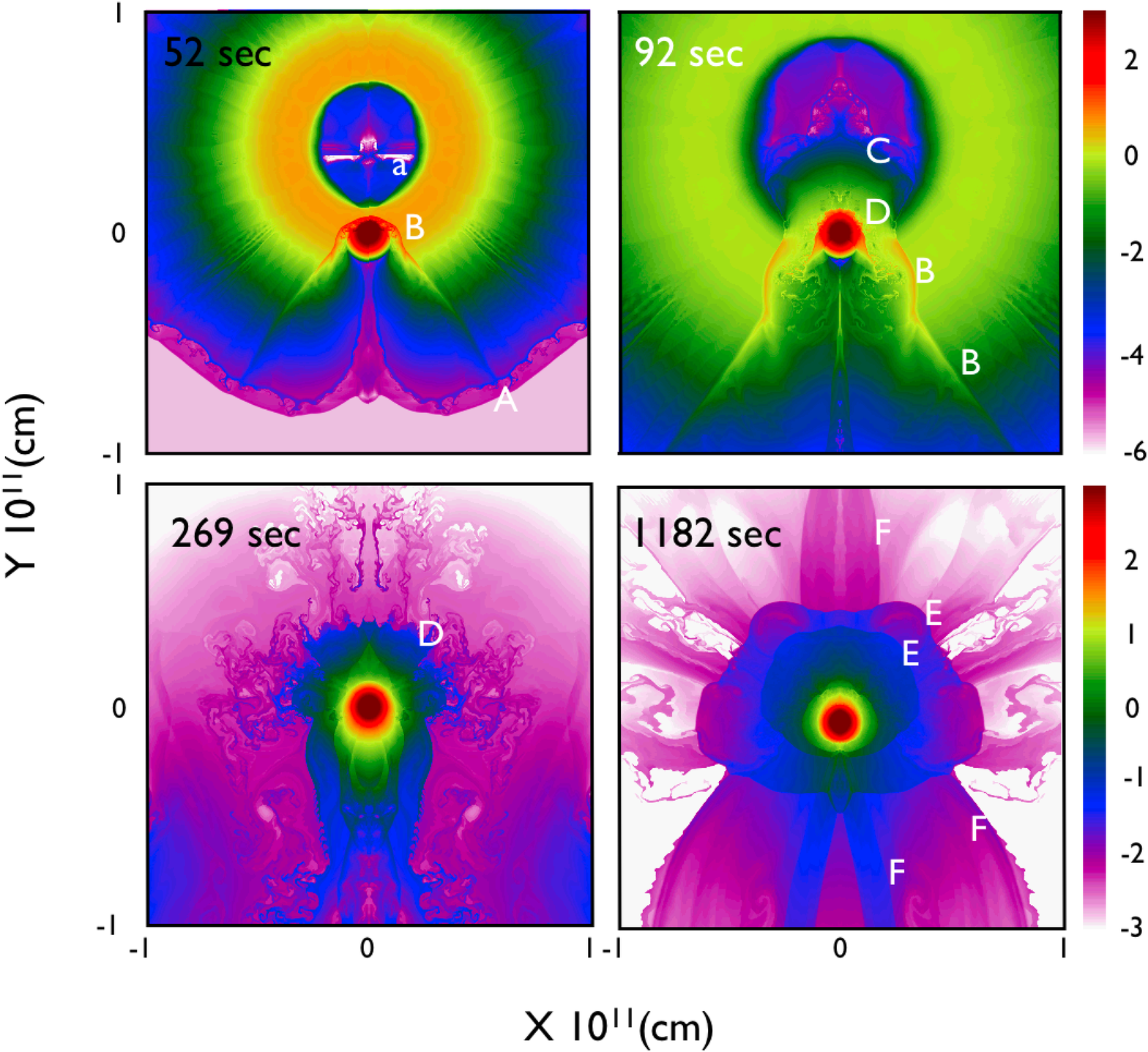}
\caption{The density distribution for model He-WDc (see Table~\ref{tab1}) with initial binary 
separation of $4 \times 10^{10}$~cm. Each frame shows a portion of the domain spanning $10^{11}$~cm. Letters refer to features described in the text. The color scale indicates the logarithm of the gas density in g~cm$^{-3}$.  \label{densS}}
\end{figure*}

\subsection{Qualitative description of evolution}
Immediately after the onset of the supernova explosion, a double shock structure is formed as the 
ejecta interact with the surrounding medium (label A in Figure~\ref{densS}; the figure illustrates the evolution of a simulation with initial binary separation of $4 \times 10^{10}$~cm, $\sim3~R_{\rm He}$, 
and 9/11 levels of refinement). 
A forward shock expands outward into the ambient medium, and a reverse shock propagates inward (in Lagrangian coordinates).
The two shocks are separated by a contact discontinuity \citep{Dwarkadas:1998p24}. During this free
expansion phase, the swept-up ambient medium has very little effect, except for the development 
of an instability at the contact discontinuity. This instability is unimportant in affecting the 
helium star, as it is seeded by a numerical grid effect and is Rayleigh-Taylor unstable (label A) 
\citep{Dwarkadas:2000p42}. The expansion of the ejecta leaves behind an extremely low-density region
at the center of the supernova (label a) that is prone to 
numerical grid effects which become visible due to the color scale; these are smoothed out later by the 
reflected shock (label C).
The SN ejecta reach the companion at $t \sim a/v_{\rm sn} \sim 50$~seconds, at which time a bow shock forms
at the leading surface of the companion, making an angle of $\sim 40^\circ$ with respect to the $z$-axis.  
This can be seen in the upper-left panel of Figure~\ref{densS} (label B)\footnote{Movies are available at\\ {\scriptsize \url{http://sipapu.astro.illinois.edu/foswiki/bin/view/Main/BinarySupernovae}}}.  
As the bow shock propagates further, the shearing of gas in conjunction with the action of gravity due to the helium star
causes distortions in the bow shock structure.  After the impact of the ejecta on the helium star, 
ejecta material begins to reflect and refill the central supernova region (label C in Figure~\ref{densS}).  
As a result of the mixing beginning at $\sim 100$ seconds associated with the Kelvin-Helmholtz and Rayleigh-Taylor instabilities in
the envelope of the helium star (label D in Figure~\ref{densS}), the bow shock breaks up and is divided into a curved shock and a straight shock that makes an angle of $\sim 40^\circ$ with the $z$-axis 
(label B in the upper-right panel of Figure~\ref{densS}). 
The instabilities continue to develop between $\sim100$~s and $\sim 600$~s
as the mixing region moves away from the helium-star envelope (lower-left panel of Figure~\ref{densS}).  
At around 100~s, the shock propagates to the center of the helium star, leading to the pulsation of the helium star.
Smooth shocks (label E in Figure~\ref{densS}) are continually generated by the oscillation of the helium star after $\sim 600$~s.
Radial shocks (label F in Figure~\ref{densS}),
which result from the interaction of shocks on the rear side of the helium star, sweep around the helium 
star (lower-right panel of Figure~\ref{densS}). As the helium star evolves further, its density profile becomes
smoother, approaching a new equilibrium state.  Qualitatively, our results are similar 
to results reported by \cite{Marietta:2000p112} and \cite{Pakmor:2008p139} for the main-sequence 
case, but with a more compact companion and smaller binary separation.  As a result of the asymmetric 
interaction, a small kick velocity, $v_{\rm kick}\sim 85$~km~s$^{-1}$, is imparted to the helium star.

\subsection{Parameter survey \label{parameters}}
We conducted a parameter survey
to explore the dependence of the numerical results on the binary progenitor's properties.
Note that \cite{Wang:2009p3585} determined the orbital period, and thus the orbital separation,
at the onset of the supernova explosion, for each choice of helium star and white dwarf.
In our study we do not follow the binary evolution up to the explosion, so the Roche-lobe radii
of Wang et al.'s models are actually larger than the radii of our helium-star models. Therefore,
in addition to varying the mass of the helium star, we also vary the binary separation in order
to determine the effect of this parameter.

\begin{figure}
\plotone{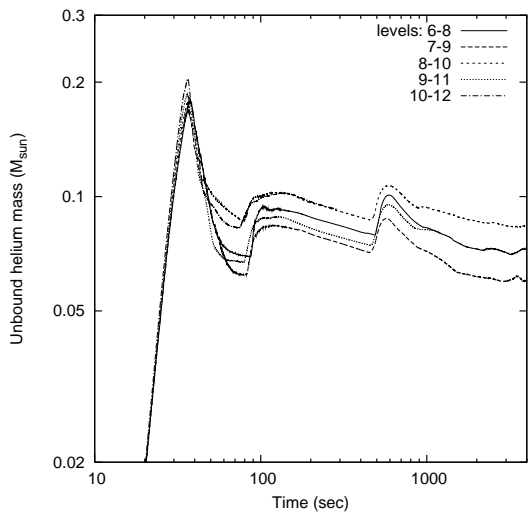}\caption{\label{convergingTest}  
Unbound helium mass versus simulation time using different levels of refinement for the He-WDc model 
(see Table~\ref{tab1}) with initial separation $3 \times 10^{10}$~cm. (Note: the 10/12 run was very
expensive and was carried out only up to 292~s.)}
\end{figure}

Figure~\ref{convergingTest} shows the typical time evolution of the amount of mass removed from the
helium star (defined as total unbound helium mass) by the SN ejecta. 
The first peak occurs at the initial impact ($\sim 30$~s), but then some gas becomes bound again
after the ejecta pass through the helium star. At $\sim 80 - 100$~seconds, a second 
peak develops which is associated with the effect of the reverse shock. Lastly, a third peak at 
$\sim 500$ seconds occurs when the compressed helium star relaxes and starts to oscillate. 
After 1,500~seconds, the amount of unbound mass reaches an approximately steady value.
The final unbound mass is thus calculated for a given initial binary separation and helium-star model
by averaging the unbound mass values computed for several time steps at this late stage. 

\begin{figure}
\plotone{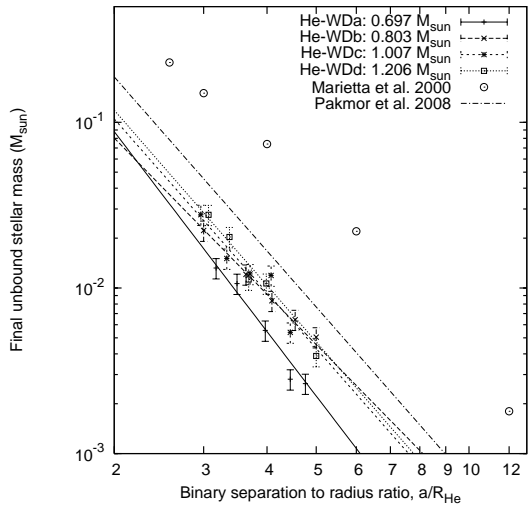}
\caption{\label{parmMass} Final unbound stellar mass versus binary separation for different helium-star models.
The separation is expressed in units of the helium-star radius. The lines show power-law relations from numerical simulations for different helium-star models (Table~\ref{tab1}). Error bars are based on the 14\% error determined in our convergence test. The dash-dot line is the fitted curve provided by \cite{Pakmor:2008p139} and the circles are the data from \cite{Marietta:2000p112}. }
\end{figure}

\begin{figure}
\plotone{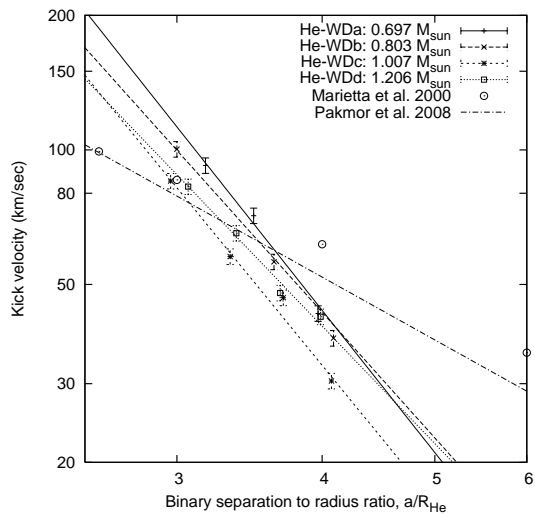}
\caption{\label{parmKick} Similar to Figure~\ref{parmMass} but for kick velocity versus binary separation for different helium-star models. Error bars are based on the 4\% error determined in our convergence test.}
\end{figure}

Figures~\ref{parmMass} and \ref{parmKick} show the final unbound mass
and helium-star kick velocity (defined as the center-of-mass velocity of the final bound 
helium) as functions of the binary separation in our simulations. We include 14\% error bars in 
Figure~\ref{parmMass} and Figure~\ref{parmKick} based on the 
results of our convergence test (\S~\ref{convergence}).
We find that the unbound mass can be fit by the relation
\begin{equation}
\delta M_{\rm ub}= C_{\rm ub} a^{m_{\rm ub}}\  M_\odot\ , \label{eq_stripped}
\end{equation}
where $a$ is the orbital separation, $m_{\rm ub}$ is the power-law index, and the constant $C_{\rm ub}$ depends only on the
helium-star model (see Table~\ref{tab1}). 
For comparison, we also plot the power-law relation with index $-3.49$ found by 
\cite{Pakmor:2008p139} and the data from \cite{Marietta:2000p112} for main-sequence 
companions (consistent with an index of $-3.14$).  
The power-law indices for our helium-star companions vary in a small range and
bracket their results, suggesting that the index may be insensitive to the evolutionary state 
of the companion. 
The normalization of the above relation does appear to be sensitive to the nature of the 
companion star.

The kick velocity of the helium star is calculated by differencing the center-of-mass positions in the 
$z$-direction at different timesteps. Alternatively, it can be determined by dividing the total bound 
helium star momentum by the total bound helium star mass.  Both methods yield the same kick velocity.
To smooth out short-term fluctuations, we average the center-of-mass positions for every ten steps,
then determine the kick velocity by numerical differentiation of these averaged values.
The helium star is initially accelerated by the SN ejecta with an acceleration $\sim G M_{\rm wd}/a^2$ for a time $\sim a/v_{\rm sn}$, and then it is kicked by the SN ejecta during the initial impact.
During the initial impact, the kick velocity varies 
dramatically. However, after $\sim 1,000$~seconds, these variations decay. Thus, we define 
the kick velocity by using the difference between the final averaged velocity and the maximum 
velocity just before the impact of the SN ejecta (the exact time range to average the final velocity varies 
from run to run and covers the entire period of ``smooth'' variation in the kick velocity). 
Figure~\ref{parmKick} shows the
kick velocity for the different helium-star models and initial binary separations.
For initial binary separations larger than 4 $R_{\rm He}$ the kick velocity could not be adequately determined, 
because the perturbed velocity from the SN is larger than the 
kick velocity at larger separations.
As obtained by
\cite{Pakmor:2008p139} and \cite{Meng:2007p73}, a power-law relation is also found in our simulation and can be fitted by the relation
\begin{equation}
v_{\rm kick}= C_{\rm kick} a^{m_{\rm kick}}, \label{eq_kick}
\end{equation}
where $v_{\rm kick}$ is the kick velocity, $m_{\rm kick}$ is the power-law index, and the constant $C_{\rm kick}$ depends only on the
helium-star model (see Table~\ref{tab1}).
However, unlike the situation for the final unbound mass, 
the slope is very different from that found in the case of main-sequence companions ($m_{\rm kick}=-1.45$ in \cite{Pakmor:2008p139}, and $m_{\rm kick}=-1.26$ in \cite{Marietta:2000p112}) (see \S4.1).

\subsection{Convergence test}
\label{convergence}
\begin{figure*}
\epsscale{1.0}
\plotone{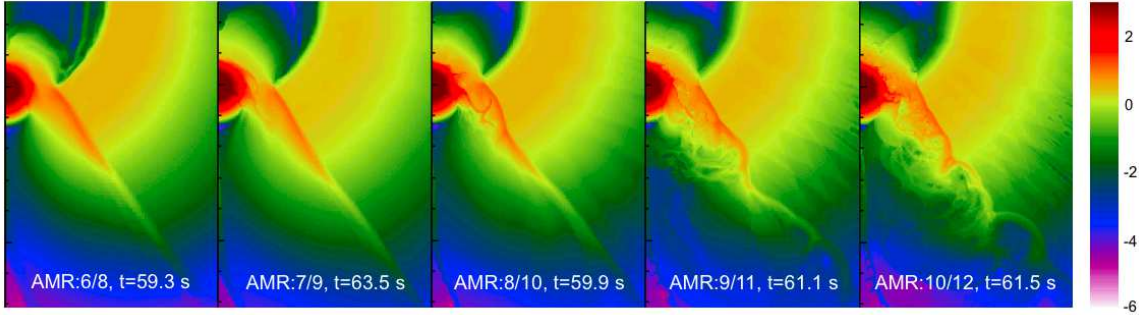}
\caption{\label{diff_amr} The density distribution for model He-WDc with initial binary separation $3 \times 10^{10}$~cm. 
Each frame shows the evolution time at around one minute for different AMR levels. The color scale indicates the logarithm 
of the gas density in g~cm$^{-3}$.}
\end{figure*}

In order to determine the robustness of our numerical results, we performed a convergence 
test for model He-WDc. We carried out simulations with the same initial binary 
separation ($a= 3\times 10^{10}$~cm) but several different maximum AMR levels, computing
the amount of unbound mass in each run as a function of time. 
The unbound mass is 
found by calculating the difference between the initial helium-star mass and the measured
total bound helium mass at each step.
The total bound helium mass is the sum of the helium masses in all zones for which the total energy is negative.
The results, which are plotted in Figure~\ref{convergingTest}, 
show that the final unbound mass lies in the range between the 8/10 and 10/12 runs.
The unbound mass is sensitive to the level of turbulence that occurs near the surface of the
helium star, particularly for resolutions higher than 8/10 (see Fig.~\ref{diff_amr}), so rather than observing
normal convergence behavior, we find that the unbound mass for runs with different resolutions fluctuates within
a small range of values.
Similar behavior is also found with other initial binary separations. 
Thus, the difference in unbound helium mass between the 8/10 and 9/11 calculations ($\sim 0.01\ M_\sun$) 
is used to estimate a relative error. Because the unbound helium mass reaches an approximately steady value at late times, this
difference suggests an estimate of about $\pm 14\%$ for the relative error in our runs, since it is not clear 
whether the 7/9 run is always the lower limit of unbound mass for different models or different separations. 
Similar analysis gives a $\pm 4\%$ relative error in kick velocity.
To allow for a feasible parameter study,
a resolution of 7/9 levels was chosen for the whole suite of twenty runs.

\section{Discussion  \label{sec_disc}}

\subsection{Stripped and ablated mass}

\begin{figure}
\plotone{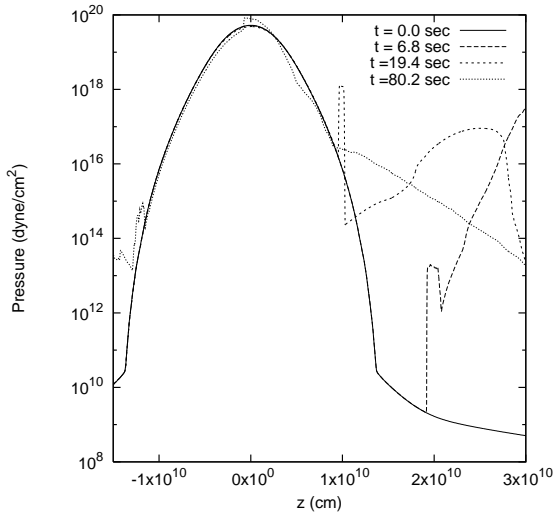}
\caption{Pressure profile along the $z$-axis at different times for model He-WDc with an initial binary separation of 
$4 \times 10^{10}$~cm and 9/11 levels of refinement. The solid line is the original pressure profile before the SN explosion. \label{1dpres}}
\end{figure}

The impact of SN~Ia ejecta on a helium-star companion is not as dramatic as for a main-sequence or red-giant 
companion like those considered by \cite{Marietta:2000p112} because helium-star companions are more compact.
This can be seen in Figure~\ref{1dpres}, which shows the gas pressure profile along the $z$-axis at several
different times during the evolution of model He-WDc with an initial binary separation of $4 \times 10^{10}$~cm.
While the pressure immediately behind the ejecta shock front is initially $\sim 10^{21}$~dyne~cm$^{-2}$, by
the time the front reaches the helium star it has dropped to $\sim 10^{13}$~dyne~cm$^{-2}$. The
shock is considerably weakened by the time it reaches the deep interior of the star. As a result,
the amount of unbound mass is much lower than in the case for main-sequence companions for a given ratio of 
separation to radius, $a/R \sim 3$, even though the helium-star channel is characterized by smaller binary 
separations.  

If we assume that the WD accretion and subsequent explosion in our models are driven by Roche lobe overflow (RLOF),
the physically appropriate initial binary separations for models $a$ to $d$ should be 3.11, 3.00, 2.84, and
2.72 times $R_{\rm He}$. The total unbound helium masses for these models at these radii (Figure~\ref{parmMass})
are $\sim 1 - 3\%$ of the initial helium-star masses, which is consistent with the suggested upper limit of $\sim 0.01\ M_\odot$ determined by
\cite{Leonard:2007p102} from observations. 

\begin{figure}
\includegraphics[angle=270,scale=0.43]{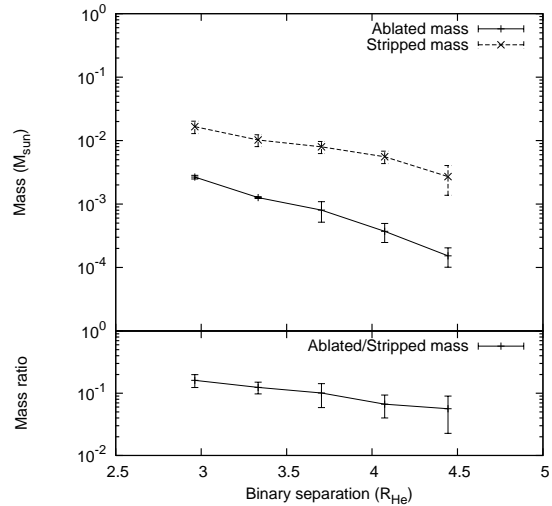}
\caption{\label{unbound} 
Comparison of averaged ablated and stripped helium mass after the initial supernova impact versus the binary separation for model He-WDc (upper panel). Error bars represent the standard deviation during the average time range. The lower panel shows the ratio of ablated to stripped helium mass.
}
\end{figure}

The unbinding of mass from the helium star results either from ablation (heating) or from stripping (momentum transfer) by the ejecta. To estimate the relative contribution of these processes, we examined the mixing of
nickel in the helium-star material. Because stripping involves the physical displacement of gas from the
helium star by the ejecta, this process is associated with the contact discontinuity between the two fluids.
This discontinuity is unstable, and the resulting instabilities lead to mixing, which while numerical in
origin, nevertheless signals contact between the nickel-rich ejecta and helium-rich stellar envelope. In
contrast, ablation proceeds through the shock heating of envelope material ahead of the contact discontinuity,
and because the shock is more stable, much less mixing is expected where ablation is dominant. 
The averaged ablated and stripped mass after the initial impact for the different cases studied is shown in Figure~\ref{unbound}.

We find that the ablated mass is comparable to the stripped mass for smaller binary separations, 
and the amount ablated is sensitive to the binary separation during the initial phase 
(less than 100 seconds for the case of model He-WDc).
The final results suggest that in most cases the amount of stripped mass can be as much as an
order of magnitude greater than the ablated mass, depending on the binary separation. 
The ratio cannot be determined accurately because we did not trace individual fluid elements in these simulations.

A simple analytical method has been used by \cite{Meng:2007p73} to estimate the amount of unbound 
matter based on conservation of momentum, ignoring the shock dynamics and the effect of ablation.  
This method yields a shallower power-law slope (-1.9) for unbound mass versus separation than 
the range that we observe (-3.1 to -4). However, 
the power-law slopes for unbound 
mass and kick velocity versus separation found by \cite{Meng:2007p73} are similar to the slopes inferred from the 
hydrogen-rich stars studied by \cite{Marietta:2000p112}. 
Although the method adopted by \cite{Meng:2007p73} 
is oversimplified, their result
suggests that ablation in main-sequence binary companions may not be as important as for helium-star binary 
companions. Figure~\ref{unbound} shows that in our simulations the ablated mass corresponds to $\sim 1-20 \%$  
of the total unbound mass, with a greater fraction ablated for smaller orbital separations. Thus, the total momentum 
imparted to the helium star results from contributions by both direct impact of the SN ejecta and
shock heating,
whereas for hydrogen-rich companions the former effect is a more important contributor than the latter.
The difference in the relative contribution of these two effects for different types of companion star may explain the difference in the power-law indices for kick velocity.

\subsection{Nickel contamination}

The companion star can be contaminated by the mixing of supernova ejecta with the helium-rich material in 
its envelope, perhaps resulting in a detectable enhanced iron abundance after the nickel radioactively decays. 
Since $^{56}$Ni is used as a tracer in our model, we have determined the amount of nickel bound to the remnant 
helium star. In general, there is a tendency for a higher level of contamination for more massive helium-star
companions or for larger orbital separations. In the former case, a massive companion presents a larger 
cross section for capture of supernova ejecta, whereas in the latter case, the ram pressure from the ejecta 
is lower for larger binary separations, resulting in a reduction of the amount of contaminated matter that is
stripped off. However, the level of contamination in our simulations is highly dependent on the development of small-scale fluid instabilities
at multifluid interfaces. Although we find little evidence for a simple relationship between the 
amount of contamination and the nature of the helium-star companion or the orbital separation, we can
estimate an upper 
limit on the nickel captured by the helium-rich companion by identifying the ejecta which cannot escape the 
gravitational potential. This limit is $\sim 5\times 10^{-4}\ M_\odot$, somewhat smaller than the estimate by
Marietta et al.\ ($\sim 1.5 \times 10^{-3}\ M_\sun$).
If the mixing of nickel is only restricted to the envelope, the nickel to helium ratio can 
be estimated by using the upper limit of nickel contamination and the final envelope mass. 
We define the core mass for the initial helium-star models by finding the radius at which the second
derivative of the density with respect to radius is a maximum; the envelope mass is then the total mass less the core mass.
Using this definition, we find that the unbound helium-star material is taken entirely from the envelope.
The final 
envelope mass can thus be obtained from the difference between the initial envelope mass and the final unbound 
mass in Figure~\ref{parmMass} at the separation corresponding to RLOF. We find that the ratio of the upper limit on bound nickel mass to the final envelope helium mass is $\sim 9-50 \times 10^{-4}$ for the different helium-star models. 
This value is substantially higher than the solar ratio of iron abundance to that of hydrogen plus helium ($5.1 \times 10^{-4}$) found by \cite{Anders:1989p4048}, suggesting that the abundance of nickel/iron in the remnant helium-star atmosphere should be enhanced relative to normal Population~I stars if surface convection is not an important factor. 

\subsection{Detecting the remnant companion star}

\begin{figure}
\plotone{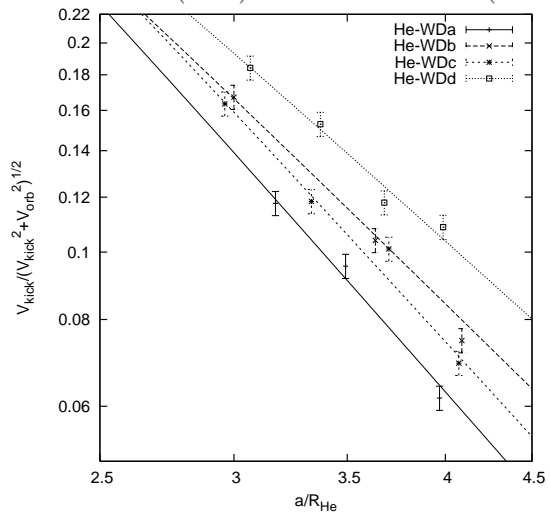}
\caption{\label{vratio} The ratio of kick speed to total speed for different helium-star models as a function of initial separation. Lines: from the fitted kick velocity. Data: from the calculated kick velocity. Error bars are based on the 4\% error determined in our convergence test. We assume the kick velocity to be perpendicular to the orbital velocity.}
\end{figure}

After the supernova,
the companion star moves with a velocity $\vec{v} \sim \vec{v}_{\rm orb}+\vec{v}_{\rm kick}$. The 
orbital speeds corresponding to the range of binary separations in our simulations are
$v_{\rm orb} \sim 350-800 $~km~s$^{-1}$, while the kick speed $v_{\rm kick} \sim 30-100$~km~s$^{-1}$. Thus, the net velocity is primarily determined by the 
orbital motion for most binary separations. Figure~\ref{vratio} shows the ratio of kick speed to the total 
speed, assuming that the kick velocity is perpendicular to the orbital velocity. If the mass transfer in the 
binary system is via RLOF, then the kick contributes little to the total velocity. For very small orbital
separations, the kick contributes at most $\sim~10 - 20\%$ of the total. 

Based on these observations,
we may expect that helium stars with high space motion found to be associated with Type~Ia supernova remnants
would be evidence for SNe~Ia proceeding through the helium-star single-degenerate channel. If such objects are
also found to have enhanced iron abundances relative to helium, this evidence would be considerably strengthened.
In such cases the relative velocity between the helium star and the centroid of the supernova remnant would
place a stronger constraint on the initial binary separation than on the initial helium star's mass or the
white dwarf mass.
 
\section{Conclusions}
We have investigated the impact of SN~Ia ejecta on a companion star in the single-degenerate helium-star channel for the short-delay-time population of SN~Ia via two-dimensional hydrodynamical simulations using a range 
of helium-star models and binary orbital separations. Although the general behavior of the temporal evolution 
is similar to that in previous studies by \cite{Marietta:2000p112} and \cite{Pakmor:2008p139} for main-sequence and red-giant companions, the amount of matter unbound from the helium stars is less than for 
hydrogen-rich companions.  Due to the shorter orbital periods of helium-star progenitor systems, the space motion of 
the companion star after the explosion is found to be higher.  We find a power-law relation between the unbound mass and initial binary 
separation that is consistent with the previous studies, suggesting that the power-law behavior is not strongly 
sensitive to the nature of the companion. The kick velocity can also fitted by a power-law and we conclude that the power-law index may reflect the relative importance of the effect of ablation.
An upper limit on the amount of nickel captured by the helium star is found to be $\sim 5\times 10^{-4}\ M_\odot$.
The ratio of nickel to helium abundance may be useful as a diagnostic of such events in future observational studies of SN~Ia stellar remnants. 

Future work in this area will include relaxing the assumption of axisymmetry to model the mixing of ejecta with 
the helium star by including the binary orbital motion in three spatial dimensions.  Additionally, including radiation transfer within these simulations
will allow us to determine how much of the helium is ionized, allowing us to make direct contact with
spectroscopic constraints on the presence of helium in these systems.
 
\acknowledgments
We thank the anonymous referee for his/her valuable comments and suggestions.
The simulations presented here were carried out using the NSF Teragrid's Ranger system at the 
Texas Advanced Computing Center under allocation TG-AST040034N.  FLASH was developed largely by 
the DOE-supported ASC/Alliances Center for Astrophysical Thermonuclear Flashes at the University of 
Chicago.  This work was supported, in part, by NSF AST-0703950 to Northwestern University. 

\bibliography{ms}


\end{document}